\documentclass[aip,adv,amsmath,amsymb,reprint]{revtex4-1}
\usepackage{graphicx}
\usepackage{bm}

\begin{document}
	\title{Design of two-dimensional particle assemblies using isotropic pair interactions with an attractive well}
	\author{William D. Pi\~{n}eros} 
	\affiliation{Department of Chemistry, University of Texas at Austin, Austin, TX 78712, USA}
	\author{Ryan B. Jadrich}
	\affiliation{McKetta Department of Chemical Engineering, University of Texas at Austin, Austin, TX 78712, USA}
	\author{Thomas M. Truskett \footnote{Corresponding author: truskett@che.utexas.edu} } 
	\affiliation{McKetta Department of Chemical Engineering, University of Texas at Austin, Austin, TX 78712, USA}
	\affiliation{Department of Physics, University of Texas at Austin, Austin, TX 78712, USA}
	
	\date{\today}
	\begin{abstract}
	Using ground-state and relative-entropy based inverse design strategies, isotropic interactions with an attractive well are determined to stabilize and promote assembly of particles into two-dimensional square, honeycomb, and kagome lattices. The design rules inferred from these results are discussed and validated in the discovery of interactions that favor assembly of the highly open truncated-square and truncated-hexagonal lattices. 
	\end{abstract}

	\maketitle 

\section{Introduction}
	The manufacture of functional materials with specific nanoscale structural features presents considerable scientific and engineering challenges. While top-down fabrication approaches (e.g., lithography) have been advanced to address such challenges, they are often too expensive or time consuming for adoption in industrial manufacturing applications.\cite{PhotonicMatsDesign,PhotonicMatsDesign2,PhotonicMatsDesign3} Bottom-up approaches such as self-assembly, on the other hand, stand as promising alternatives in which material building blocks (nanoparticles, block copolymers, etc.) might be designed--through modification of their effective mutual interactions \cite{ColloidInteractionTuning_1,ColloidInteractionTuning_2,SelfAssemblyForcesReview,ColloidInteractionsRev}--to spontaneously self-organize into a state that exhibits the desired microstructural features.\cite{SelfAssemblyPolyhedraParticles,GeometricParticlesAssembly,SelfAssemblyPatchyParticles1,JanusParticlesSelfAssemblyRev} To determine which interactions stabilize a desired self-assembled structure, a design strategy is needed. Traditional `forward' approaches discover promising material combinations through screens (i.e., combinatorial searches) or more limited testing of candidate systems judiciously chosen based on physical intuition.  Alternatively, `inverse' design strategies provide a more direct means for discovering interactions suitable for stabilizing the target structure, typically via solution of a constrained optimization problem.  \cite{InvDesignTechRev,InvDesignGeneral,InvDesignEntropy,InvDesignPerspective}

	A classic example of an inverse design problem for materials is the discovery of an isotropic pair potential $\phi(r;\{\alpha\})$ between particles that stabilizes a specified crystalline lattice arrangement as the ground state (GS). Here, $r$ is the distance between particle centers and $\{\alpha\}$ is the set of optimizable parameters that, along with the specified functional form, define the pair potential. Researchers have developed robust inverse design strategies that have been applied to discover isotropic interactions that stabilize a variety of open structures in two and three dimensions under various constraints (including honeycomb\cite{MT_SquareHoneyConvexCom,MT_SquareHoneyConvexFull,RT_HoneyDoubleWell,AvniDimTransfer}, kagome\cite{ZT_MuOptKagomeAsymLats,InvDesignKagome,InvDesignKagomeFunctionalMethod}, simple cubic\cite{Avni3DLattices,AvniDimTransfer}, and diamond\cite{InvDesignKagomeDiamond,Avni3DLattices,MT_DiamondConvex} lattices, to mention a few).  Recently, a reformulation of this type of GS optimization problem was introduced\cite{SquLat_dmu_opt} which significantly improved performance and allowed for (1) exploration of how design goals affect trade offs for the target phase (e.g., thermal versus volumetric stability \cite{SquLat_dmu_opt}) and (2) discovery of interactions that stabilize very challenging target structures (e.g., snub square\cite{KagSnub_gams_opt}, truncated square and truncated hexagonal\cite{Truncs_gs_opt} lattices). An advantage of GS-focused optimization is that, because it requires the {\em a priori} identification of structures that most closely compete with the target lattice, it can offer insights into how those competitions influence the functional form of the optimized interactions. On the other hand, the identification of such competitors can be highly nontrivial, especially for target lattices with open structures. Furthermore, stabilization of the target as the GS does not guarantee that the structure will be kinetically accessible via a self-assembly process at higher temperature. 
	
	An alternative, simulation-based design method focused on relative-entropy (RE) maximization was recently shown\cite{RelEntropy_2D_structures,RelEntropy_general} capable of discovering interactions that promote {\em spontaneous assembly} of a target lattice from the disordered fluid upon cooling. Notably, the RE optimization computes interactions ``on-the-fly'' based on structures accessed in a simulation and thus avoids the need to explicitly identify possible competing structures in advance. As a result, the RE approach has been able to help design isotropic interactions that favor unusually open ordered phases (e.g., truncated hexagonal \cite{RelEntropy_2D_structures} and truncated tri-hexagonal \cite{RelEntropy_general} lattices, which naturally compete with a variety of stripe-phase structures that are difficult to identify {\em a priori})  as well as disordered hierarchical structures (e.g., porous mesophases\cite{inv_design_porous_mesophases,RelEntropy_general} and cluster fluids\cite{inv_design_clusters,RelEntropy_general}, which are not easily designable with GS-based optimization strategies). However, because RE maximization does not require identification of competing structures, it cannot provide direct insights into how the optimized interactions stabilize the target relative to its competitors.     
	
	Here, we use both GS and RE optimization strategies to investigate a thus far unanswered question: which open two-dimensional crystal structures can be stabilized by an isotropic pair interaction comprising a repulsive core and a single attractive well? Interactions of this form are ubiquitous in nature, e.g., those present in noble gases and liquid metals\cite{effective_interaction_liquid_metal}, the former of which have thermodynamic properties that are well captured by the familiar Lennard-Jones model. Single-well (effective) interactions can also arise in colloidal systems where, e.g., depletion interactions, van der Waals forces, and screened Coulomb interactions are present and can be tuned via the material selections made for the particle core, surface-passivating polymers or ligands, co-solutes, and the solvent.\cite{ColloidInteractionTuning_2,DNA_nanoparticle_single_well,DNA_nanoparticle_control,single_well_colloid_polymer_mix} Given the diverse contexts in which single-well interactions naturally emerge and can even be systematically modified, it is of interest to theoretically explore how the combination of a repulsive core and an attractive well might be chosen to favor various targeted crystal structures.

	Our specific approach is to study the inverse design of isotropic, single-well pair potentials to stabilize 2D crystal structures (square, honeycomb, kagome, truncated square, and truncated hexagonal) using GS and RE methodologies described in section~\ref{sec:Methods}. In section~\ref{sec:Results}, we present our results in the context of `design rules' that can be inferred from our study of square, honeycomb, and kagome lattices.  We then apply these design rules to discover single-well interactions that stabilize the more challenging truncated square and truncated hexagonal lattices.  We conclude with section \ref{sec:Conclusion}, where we summarize the rules and provide some further thoughts about their application to the inverse crystal design process. 
 
\section{Methods} 
\label{sec:Methods}
\subsection{Ground State Optimization}
\subsubsection{Analytical Formulation}  
    In this method, we seek an isotropic, single-well interparticle pair potential $\phi(r;\{\alpha\})$ that stabilizes a target lattice $l_t$ as the ground state over (at least) a narrow density range. Broadly speaking, we formulate the design as an analytical nonlinear program and determine the pair potential parameters $\{\alpha\}$ via its numerical solution using GAMS (General Algebraic Modeling System)\cite{GAMSWorldBank,GamsGuide2013,GamsSoftware2013}.
    
	For the model pair interaction, we choose the functional form  $\phi(r;\{\alpha\})$, 
	\begin{equation}
	\phi(r)/\epsilon=
	\begin{cases} 
		A/r^{n} \exp{(-r^2/\sigma_0)}+\sum^{2}_{i} B_i \exp{(-(r-r_i)^2/\sigma_i)} + f_{\text{shift}}(r) & r < r_{\text{c}}  \\
		0 	& r \geq r_{\text{c}} 
		\label{eq:potential} 
	\end{cases}
	\end{equation}
where $r=r/\sigma$, $A,n,\sigma_0,B_i,r_i$ and $\sigma_i$ are design parameters (i.e. $\{\alpha\}$), $r_c$ is the cut-off radius and $f_{\text{shift}}(r)= P r^2 + Q r + R$ is a quadratic shift function added to enforce $\phi(r_{\text{c}})= \phi'(r_{\text{c}})= \phi''(r_{\text{c}})= 0$.  
Location of the potential minimum $r_{\text{min}}$ is fixed as 
	\begin{equation}
		\phi'(r_{\text{min}}) = 0	
	\end{equation}
and an appropriate well profile is achieved by means of the following constraint equation 
	\begin{equation}
	\begin{aligned}
		\phi'(\mathbf{r}_l) &< 0	& r < r_{\text{min}} 	\\
		\phi'(\mathbf{r}_r) &> 0	& r > r_{\text{min}} 	
	\end{aligned}
	\end{equation}
where $\mathbf{r}_i$ denotes an appropriately distributed set of radial points left($i=l$) or right($i=r$) of $r_{\text{min}}$. The magnitude of the minimum was fixed such that $\phi(r_{\text{min}}) = -\epsilon$. Finally, we constrain $B_1$ and $\sigma_0$, $\sigma_2$ as 
	\begin{equation}
	\begin{aligned}
		B_1 	 &< 0 	\\	
		\sigma_1 &> 2 \sigma_0 \\
		\sigma_1 &> \sigma_2 
	\end{aligned}
	\end{equation}
To simplify notation, quantities are implicitly nondimensionalized from this point forward in terms of the usual (dimensionally appropriate) combination of parameters, including the energy scale $\epsilon$, the length scale $\sigma$, or the Boltzmann constant $k_B$.

	We compute quantities of interest over a narrow density range using the canonical ensemble. In particular, we consider a grid of densities $\{\rho_j\}$ that spans a range $\Delta\rho={1.21-1.23}$ to compute the potential energy $U$ of the target lattice and equidensity competing structures,
	\begin{equation} 
		U(\rho_j)=\frac{1}{2} \sum^{r_{\text{c}}}_{r} n_i \phi(r_i;\rho_j)
	\end{equation}
where the sum is over all lattice-dependent coordination shells $n_i$ at distances $r_i$ up to the pair-potential cut-off. Similarly, the potential energy for phase-separated competitors $U_{\text{PS}}$ was computed as\cite{Shellbook}
	\begin{equation} 
		U_{\text{PS}}(\rho_j)=(1-x)(U_2-U_1)-U_1
	\end{equation} 
where $x$ is the molar fraction for the crystal phase $l_1$, $U_{\text{PS};}(\rho_j)$ is the total system energy at net density $\rho_j$ and $U_i$ is the potential energy for crystal $l_i$ at density $\rho_i$ (see supplementary material for further discussion of this equation). Due to the narrow density range, energies of phase separated structures at only the midpoint density of the range $\rho_m$ were sufficient to consider for optimization purposes. Stability of the target lattice relative to these competitors for the remainder of density points within the range was verified by means of forward ground-state phase diagram calculations, as discussed below. 

	Finally, we define an objective function $F$ that maximizes the sum of the energy differences of the target and competitors at the midpoint density $\rho_{m}=1.22$ as 
	\begin{equation}
		F = \sum_{l_i} U_{l_i}(\rho_m) - U_t(\rho_m)
	\end{equation}
where $U_t(\rho_m)$ denotes the target's energy and $U_{l_i}(\rho_m)$ the energy of a competitor lattice $l_i$. Stability of the target lattice for the remaining grid points $\{\rho_j\}$  was ensured by constraining the energy differences for every competitor $l_i$ as 
	\begin{equation}
	     U_{l_i}(\rho_j) - U_t(\rho_j) > 0 
	\end{equation}
Below, we discuss how structures that closely compete with the target are identified.

\subsubsection{Competing Structures}
 	The pool of closely competing structures for the GS optimization was determined in the following way. First, an initial competing pool comprising lattices commonly stabilized by isotropic potentials was chosen, and the optimization was performed based on comparisons of the target lattice with that pool. The resulting optimized pair potential was then used to a carry out a `forward calculation' in which the stability of the target lattice was determined relative to a comprehensive list of possible competing lattices (including those with variable parameters chosen via optimization using GAMS as well as lattices co-existing at different densities).  Any structures found to be more energetically stable than the target were then added to the competing pool for a subsequent optimization. This process was repeated until the forward calculation of the optimized pair potential yielded no new competitors and established the target as the energy minimum, a point at which the optimization procedure was considered converged and completed. The detailed calculation procedure for single phases has been explained in detail previously \cite{KagSnub_gams_opt}. A discussion of how to consider phase-separated crystals is provided in the supplementary material. 

	Following this approach, the final competing pools for square, honeycomb and kagome lattices were determined to be as follows: 
	\begin{itemize}
	\item[]	Square target: hexagonal, kagome, snub hexagonal, snub square, elongated triangular, and rectangular ($b/a=1.17$) lattices, as well as phase-separated lattice structures comprising square and hexagonal lattices which we denote by (square lattice density)/(hexagonal lattice density) and the corresponding square lattice molar fraction: (1) 1.1479/1.4344, 0.7040; (2) 1.2675/0.7008, 0.9518; (3) 1.2512/0.7095, 0.9665; (4) 1.1833/1.44961, 0.83636; (5) 1.2279/0.7050, 0.9912.
	\item[]	Honeycomb target: hexagonal, square, kagome, snub hexagonal, elongated triangular, rectangular ($b/a=1.59$) lattices, and three specialized competitors, two of which resembled a distorted honeycomb and a staggered elongated triangular (see figure 1S in the supplementary material). There was a single phase-separated structure composed of honeycomb and hexagonal crystals which we denote by (honeycomb density)/(hexagonal density) and honeycomb molar fraction: 0.49180/1.86031 0.18861. 
	\item[]	Kagome target: hexagonal, square, snub hexagonal, elongated triangular, honeycomb, kagome ($b/a=1.02$), and twisted kagome lattices, as well as a specialized competitor resembling rows of staggered rectangular boxes (see figure 2S in the supplementary material). No phase-separated competitors were found for this target.
	\end{itemize}

\subsection{Relative Entropy Optimization}
Relative entropy course graining relies on maximizing the probability of achieving a desired equilibrium configurational ensemble by optimizing a set of parameters $\boldsymbol\theta =\{ \theta_1,\theta_2,...,\theta_m \}$ of the interaction pair potential $u(r|\boldsymbol\theta)$. This can be achieved by direct (``on-the-fly'') optimization in a simulation,\cite{RelEntropy_2D_structures} where potential parameters are periodically updated as
	\begin{equation}
		\boldsymbol \theta^{i+1} = \boldsymbol \theta^{i}+ \gamma \int_{0}^{\infty} dr r[  g(r|\boldsymbol \theta^i)-g_{\text{tgt}}(r)   ]
		[ \boldsymbol \nabla_\theta u(r|\boldsymbol\theta) ]_{\boldsymbol\theta=\boldsymbol\theta^i}
	\label{eq:REupdateStep}
	\end{equation}
where $i$ indicates iteration step, $g(r|\boldsymbol\theta^i)$ is the current radial distribution function, $g_{\text{tgt}}(r)$ is the target lattice radial distribution function and $\gamma$ is an adjustable scalar parameter to ensure stability and convergence of the optimization. Detailed derivation of this update procedure is available in previous work\cite{RelEntropy_2D_structures,RelEntropy_general}. In this paper, we use $u(r|\boldsymbol\theta)$ consisting of Akima splines whose knots are the parameters. We constrain the splines to display the features of a single attractive well with a minimum located at $r_{\text{min}}$. If we consider a set of scalars $\mathbf{r}=\{r_1, r_2,..., r_{\text{m}}\}$ as denoting the position of the Akima knots, the well constraint is easier to implement by considering the parameters $\boldsymbol\theta \equiv \theta(\mathbf{r})$ as differences of the pair potential 
	\begin{equation} 
		\theta(r_k)=-[u(r_{k+1})-u(r_{k})]
	\end{equation} 
where the sign has been inverted to resemble a force term and for simulation convenience. The desired interaction form can then be achieved by constraining positions to the left ($\mathbf{r}_l$) and right ($\mathbf{r}_r$) of the minimum as
	\begin{equation}
	\begin{aligned}
		\theta(\mathbf{r}_l) &> 0	& \mathbf{r}_l < r_{\text{min}} 	\\
		\theta(\mathbf{r}_r) &< 0	& \mathbf{r}_r > r_{\text{min}} 	
	\end{aligned}
	\end{equation}
Lastly, the minimum position itself is free to vary and is updated as follows
	\begin{equation} 
	\begin{aligned}
	\text{if	}& u(r_{\text{min}})^{i+1} > 0 & \text{ and }& |u(r_{k+1})^{i+1}| < |u(r_{\text{min}})^{i+1}| &  \text{	then	}& r_{\text{min}}=r_{k+1} \\
	\text{if	}& u(r_{\text{min}})^{i+1} < 0 & \text{ and }& |u(r_{k-1})^{i+1}| < |u(r_{\text{min}})^{i+1}| &  \text{	then	}& r_{\text{min}}=r_{k-1} 
	\end{aligned}
	\end{equation}
where $|...|$ indicates absolute value, the $i+1$ superscript indicates the next iteration round, and $k$ subscripts indicates the knot position immediately to the left ($k-1$) or right ($k+1$) of the $r_{\text{min}}$ position at the $i$th iteration. Derivation of this result is provided in the supplementary material. 

\subsection{Molecular Simulations}
\subsubsection{Molecular Dynamics For Relative Entropy Optimization}
The Relative Entropy (RE) optimization is implemented with the GROMACS 4.6.5 molecular dynamics package \cite{GROMACS1,GROMACS2}. Briefly, a system of particles ($N=200-1000$) interacting via a single-well spline pair potential is simulated in the canonical ensemble using a periodically replicated rectangular simulation cell with aspect ratio chosen to accommodate the target lattice. The system is initiated in the target lattice at $T=1$, melted via a run of $2\times10^6$ time steps at $T=1.5$, and then slowly cooled back to $T=1$ (over $5\times10^6$ time steps). Radial distribution function statistics $g(r)$ are collected over the last $10^6$ time steps and compared to the target's structure $g_{\text{tgt}}(r)$. The spline potential is then updated per Eq.~\ref{eq:REupdateStep} and the process is iterated in this manner. In practice, $\gamma=0.005-0.01$ is sufficient for all crystal target structures considered here. Convergence is typically achieved in about 100-200 iterations, and optimization is considered complete once the self-assembled crystal remains stable up to $T=1.5$.  

\subsubsection{Monte Carlo Simulations}
Monte Carlo simulations were used to test for stability of target lattices of particles interacting via the potentials obtained from the ground-state optimizations. In particular, systems of $N$ particles ($N=64$, $40$, $108$ for square, honeycomb, and kagome targets, respectively) interacting via Eq.~\ref{eq:potential} with their respective ground-state optimized parameters were simulated in the canonical ensemble using a periodically replicated simulation cell with densities and aspect ratios chosen to best accommodate the target lattices. In each case, the target was melted to form a liquid at a high temperature and then either quenched (square lattice) or slowly cooled (over $\sim 5\times10^6$ Monte Carlo steps for the honeycomb and kagome lattices) to low temperature to check for assembly of the target crystal. For the square, honeycomb, and kagome lattices, the aforementioned high temperatures were $T=0.5091$, $1.2091$, and $2$ and the low temperatures were $T=0.1627$, $0.5$, and $0.5$.  For the 24 Monte Carlo simulations we performed for the three targets, we observed perfect assembly in $50\%$ (square), $17\%$ (honeycomb), and $100\%$ (kagome) of the runs.        

\section{Results and Discussion}
\label{sec:Results} 

\begin{figure}[ht]
	\includegraphics[scale=0.35]{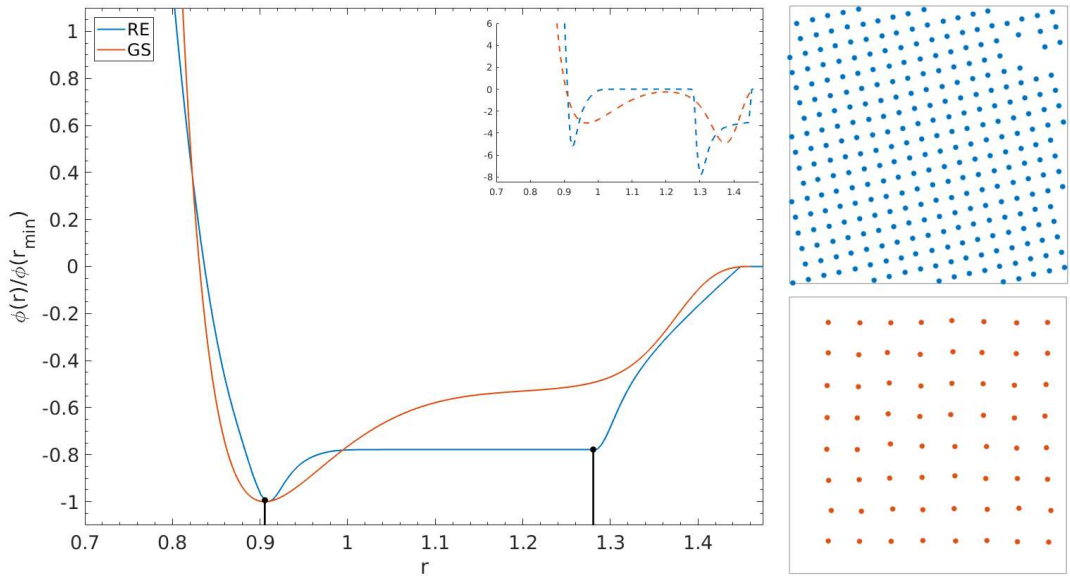}
	\caption{Left: Inversely designed pair potentials $\phi(r)/\phi(r_{min})$ (solid lines) for the square lattice from RE (blue) and GS (red) optimization strategies described in Sec.~\ref{sec:Methods}, where $r_{min}$ denotes radial position of the well minimum. Normalized forces $-\phi'(r)/\phi(r_{min})$ are shown in the inset. Black vertical lines denote the ideal coordinate shell positions of the target structure at the optimization density. Potential parameters for the GS optimization are provided in the supplementary material. Top right: Configuration from a molecular dynamics simulated annealing run using the RE optimized pair potential. Bottom right: Configuration from a Monte Carlo quench of the GS optimized pair potential.}
	\label{fig:squ_pots}
	\end{figure}

\begin{figure}[ht]
	\includegraphics[scale=0.35]{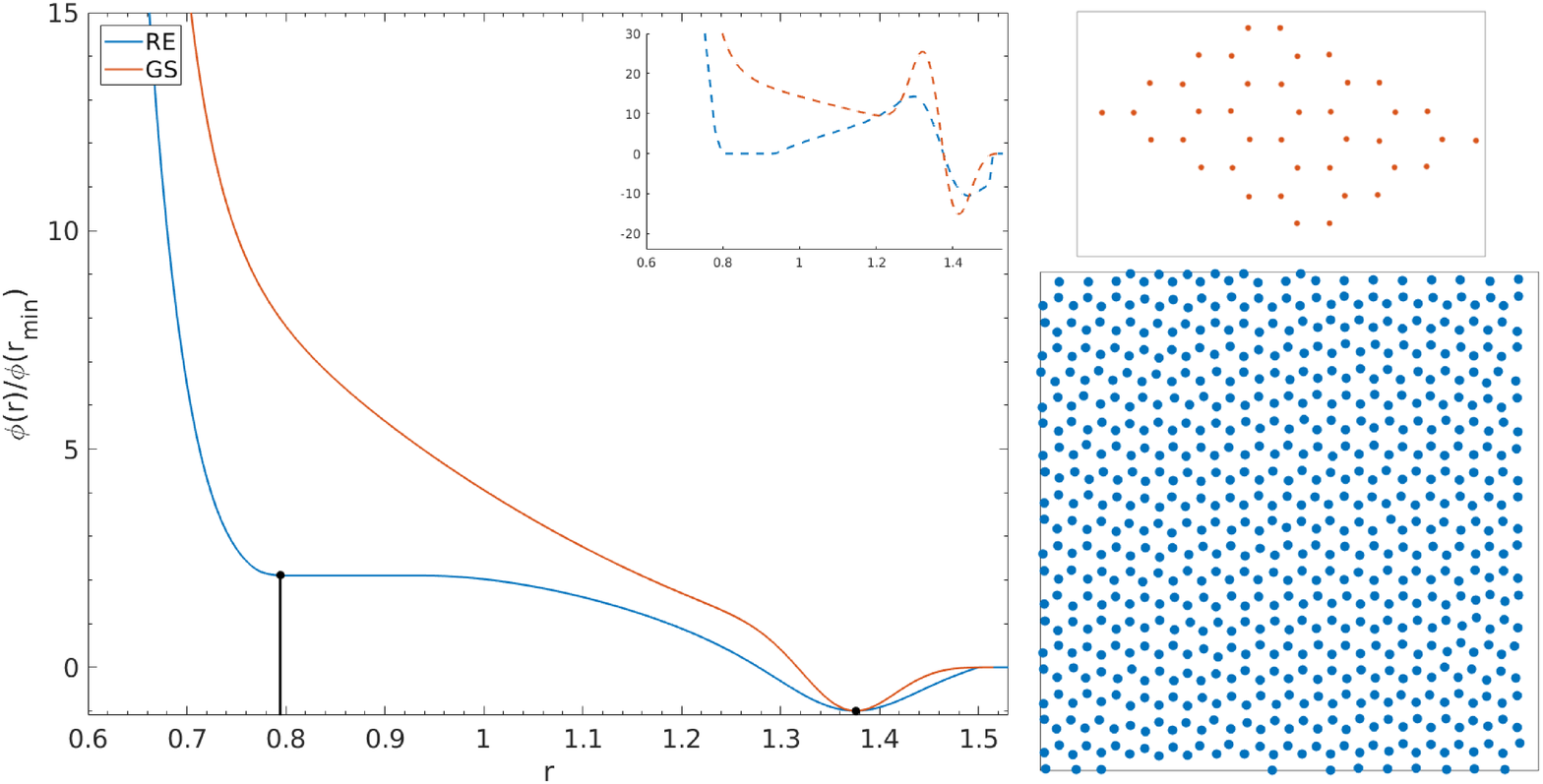} 
	\caption{Left: Inversely designed pair potentials $\phi(r)/\phi(r_{min})$ (solid lines) for the honeycomb lattice from RE (blue) and GS (red) optimization strategies described in Sec.~\ref{sec:Methods}, where $r_{min}$ denotes radial position of the well minimum. Normalized forces $-\phi'(r)/\phi(r_{min})$ are shown in the inset. Black vertical lines denote the ideal coordinate shell positions of the target structure at the optimization density. Potential parameters for the GS optimization are provided in the supplementary material. Top right: Configuration from a molecular dynamics simulated annealing run using the RE optimized pair potential. Bottom right: Configuration from a Monte Carlo quench of the GS optimized pair potential.}
	\label{fig:hc_pots}
	\end{figure}

\begin{figure}[ht]
	\includegraphics[scale=0.35]{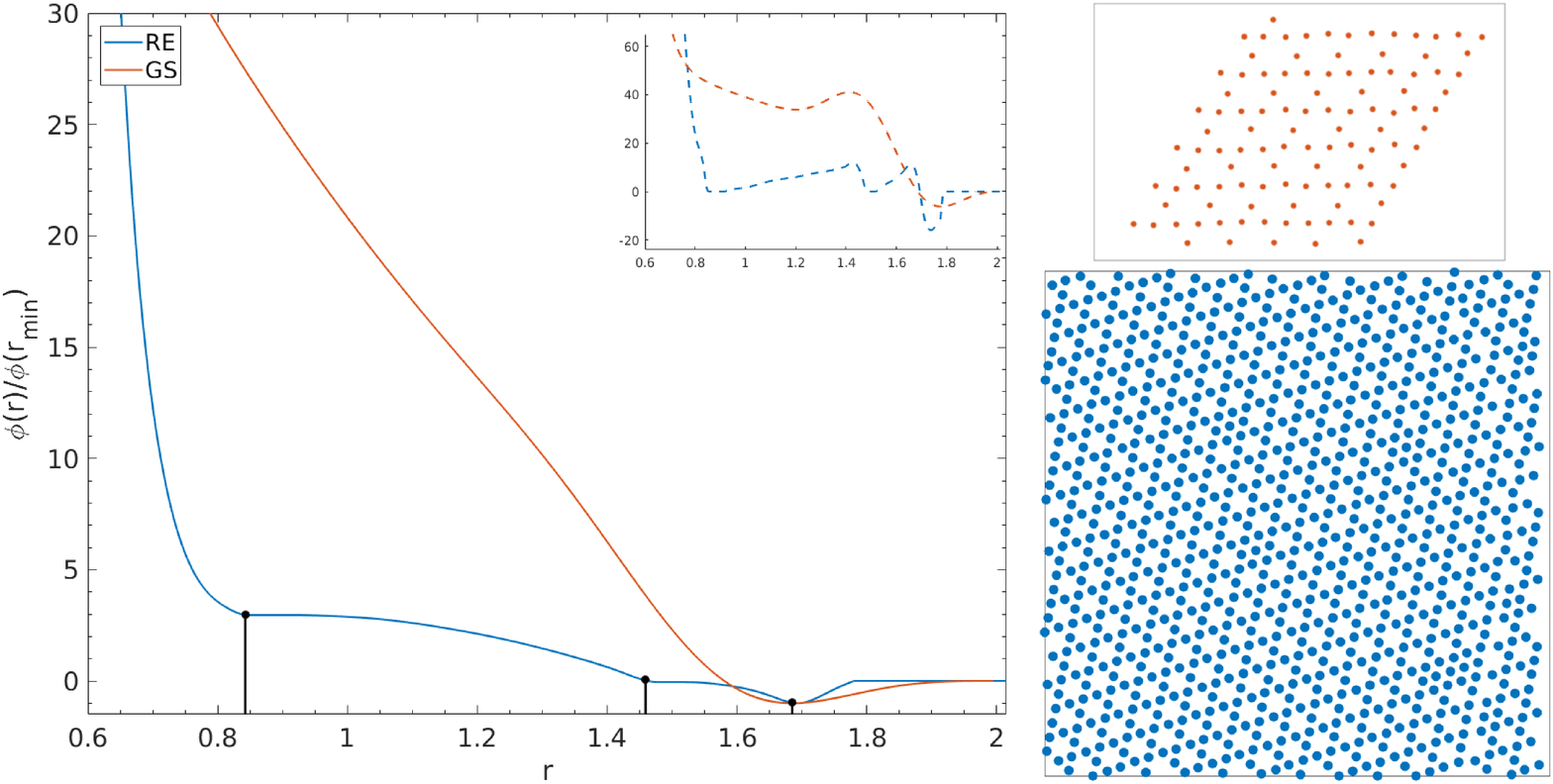} 
	\caption{Left: Inversely designed pair potentials $\phi(r)/\phi(r_{min})$ (solid lines) for the kagome lattice from RE (blue) and GS (red) optimization strategies described in Sec.~\ref{sec:Methods}, where $r_{min}$ denotes radial position of the well minimum. Normalized forces $-\phi'(r)/\phi(r_{min})$ are shown in the inset. Black vertical lines denote the ideal coordinate shell positions of the target structure at the optimization density. Potential parameters for the GS optimization are provided in the supplementary material. Top right: Configuration from a molecular dynamics simulated annealing run using the RE optimized pair potential. Bottom right: Configuration from a Monte Carlo quench of the GS optimized pair potential.} 
	\label{fig:kag_pots}
	\end{figure}

	As demonstrated below (see Fig.~\ref{fig:squ_pots}-\ref{fig:truncs_sa}), we were able to discover, using the inverse design methodologies of RE and GS optimization and the Monte Carlo simulations described in Sec.~\ref{sec:Methods}, isotropic pair potentials with a single well that promote self-assembly of all targeted lattices. Below, we examine the resulting optimized pair potentials and briefly discuss them within the context of the `design rules' that they suggest for this class of systems. 


\subsection*{1. The interaction range must span a minimum number of coordination shells} 

	This rule of thumb states the intuitive idea that the range of an isotropic interaction potential must be large enough to distinguish (and hence provide energetic advantage to) the target structure compared to its competitors. While only strictly true at zero temperature, we find that it provides a helpful guide for designing isotropic pair potentials to stabilize and promote assembly of our target lattices. Evidence of this rule's utility was recently demonstrated in an RE optimization study where the shortest range isotropic, repulsive pair potential for stabilizing each of a wide variety of two-dimensional lattices was determined.\cite{RelEntropy_2D_structures} In that work, it was shown that the square and honeycomb lattices could be stabilized by an interaction spanning only two coordination shells, while the kagome lattice required a potential spanning three shells. These results are consistent with the single-well pair potentials we obtain here from optimizations for the corresponding target lattices (see black bars denoting coordination-shell locations in figures \ref{fig:squ_pots}, \ref{fig:hc_pots}, and \ref{fig:kag_pots} for square, honeycomb and kagome lattices respectively). As discussed below, the single-well potentials designed to stabilize truncated square and truncated hexagonal lattices also span the minimum number of shells previously reported for these structures.\cite{RelEntropy_2D_structures} Consistent with this rule, our attempts to design single-well pair potentials that stabilize any of these structures using shorter range interactions were unsuccessful, and simulations employing the resulting optimized potentials led to either amorphous (glassy) or competing crystalline structures. Consistency with this rule may also be found retrospectively in the other published works featuring the same design targets (e.g., square\cite{MT_SquareHoneyConvexFull,SquLat_dmu_opt,AvniDimTransfer} and honeycomb \cite{RT_HoneyDoubleWell,MT_SquareHoneyConvexFull,AvniDimTransfer}) and was briefly hinted at in another recent inverse design study\cite{ZT_MuOptKagomeAsymLats}. Thus, while the results of the present study are not the first to suggest this design rule, they solidify the notion that a minimum interaction range must typically be spanned by an isotropic potential to favor a target structure compared to its competitors, an observation which acts as the foundation for the next two rules.  

\subsection*{2. The attractive well must be narrow, spanning only the most distant coordination shell(s) in the required range}  
	Given the observation that an isotropic pair potential requires a minimal range to stabilize a given target structure relative to its competitors, one might ask whether one can expect that a simple (e.g., Lennard-Jones-like) pair potential with a wider range of negative energy (i.e., a wider well) that spans the first and more distant coordination shells in the target lattice can stabilize any of the open structures considered here. Based on the results of our inverse design study, the answer is no, save for the square lattice. The optimized potentials for the square lattice target (see figure \ref{fig:squ_pots}) have their minimum in the first-neighbor shell, which may at first glance appear counterintuitive since the first shell of the closely competing triangular lattice has more neighbors (six versus four) and hence a more favorable first-shell contribution to the energy. However an energetic advantage of the square lattice is still accomplished with this interaction by having a well that is wide enough to reach the second neighbors of the square lattice but not the more distant second neighbors of the triangular lattice.
	
	For the more complex honeycomb and kagome structures, we found that the attractive well feature had to be constrained to the last one or two (closely spaced) shells of the interaction range (see plots in figure \ref{fig:hc_pots} and \ref{fig:kag_pots}). If the pair potential wells were allowed to incorporate nearer coordination shells as well, the optimizations and subsequent forward calculations would inevitably find that phase-separated structures involving a more highly coordinated and hence more energetically stable (typically triangular) lattice were more favorable than the single-phase target structure. This was true for either GS or RE optimization strategies, and the qualitative similarity of the designed pair potentials (and corresponding forces shown in the figure insets) points toward the generality of this observation.  
	
\subsection*{3. Nontrivial repulsive features (e.g., shoulders) are needed in addition to the attractive well in the pair potential to stabilize more complex targets} 
	Given the constraints on the location and width of the attractive well discussed in the rules above, there is still a question of how simple the repulsive profile ($r<r_{min}$) can be in a single-well pair potential and simultaneously stabilize the target structure. As can be seen from the pair potentials and forces shown in figures \ref{fig:hc_pots} and \ref{fig:kag_pots} for the honeycomb and kagome lattices, nontrivial features in the repulsive interactions naturally emerge from the GS and RE optimizations for low-coordinated target structures. As emphasized in recent inverse design studies using repulsive isotropic potentials \cite{SquLat_dmu_opt,KagSnub_gams_opt, Truncs_gs_opt}, these features are tied to the location of coordination shells in competing lattices and necessary to stabilize target structures relative to more highly coordinated lattices, including the triangular crystal, which are otherwise naturally favored by smooth, short-range repulsive interactions. Instead of shoulders, the presence of multiple positive wells in the potential (if allowed in the optimization) could also have the same effect. Importantly, the inclusion of a single well in the pair interaction does not obviate the need for shoulders or other nontrivial repulsive features in the potential for $r<r_{min}$.      
	
	Having identified the three qualitative rules above, we conclude by presenting their application to the design of the more challenging, highly open structures of the truncated square lattice and the truncated hexagonal lattice. Pair potentials stabilizing these structures with purely repulsive interactions have very recently been reported using both the RE\cite{RelEntropy_2D_structures,RelEntropy_general} and the GS\cite{Truncs_gs_opt} optimization strategies. However, due to the known complexity and tediousness involved in targeting these open structures with GS calculations (specifically, in identifying the large set of relevant competing structures which include a multiplicity of stripe phases), here we limit the pair potential discovery to the RE method. As such, following rule 1, we know in advance from the published studies the expected minimum interaction range to stabilize these targets. Next, we anticipate that the single well feature must lie towards the end of the interaction well and not be too broad as per rule 2. Finally, we expect the stabilization to require the introduction of multiple repulsive shoulders in the pair potential before the well, as per rule 3. 
	\begin{figure}[ht]
	\includegraphics[scale=0.50]{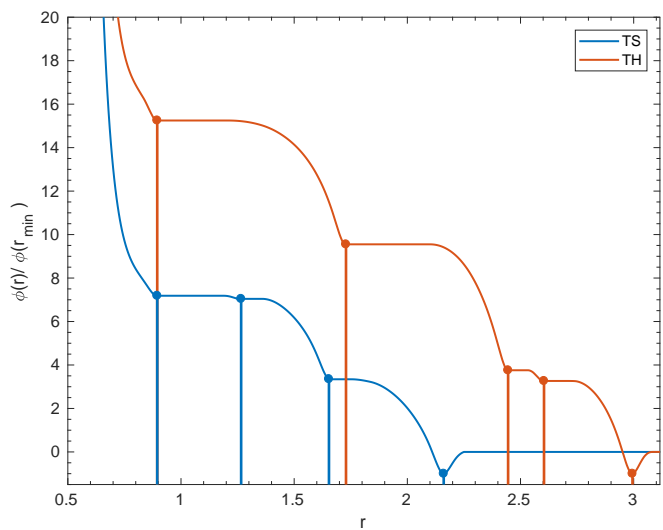} 
	\caption{Inversely designed pair potentials $\phi(r)/\phi(r_{min})$ (solid lines) for truncated square (blue) and truncated hexagonal (red) lattices using the RE optimization strategy described in Sec.~\ref{sec:Methods}, where $r_{min}$ denotes radial position of the well minimum. Vertical lines denote the respective coordinate shell positions for the targeted lattices at the optimized density.}
	\label{fig:truncs_pots}
	\end{figure} 
	\begin{figure}[ht]
	\includegraphics[scale=0.50]{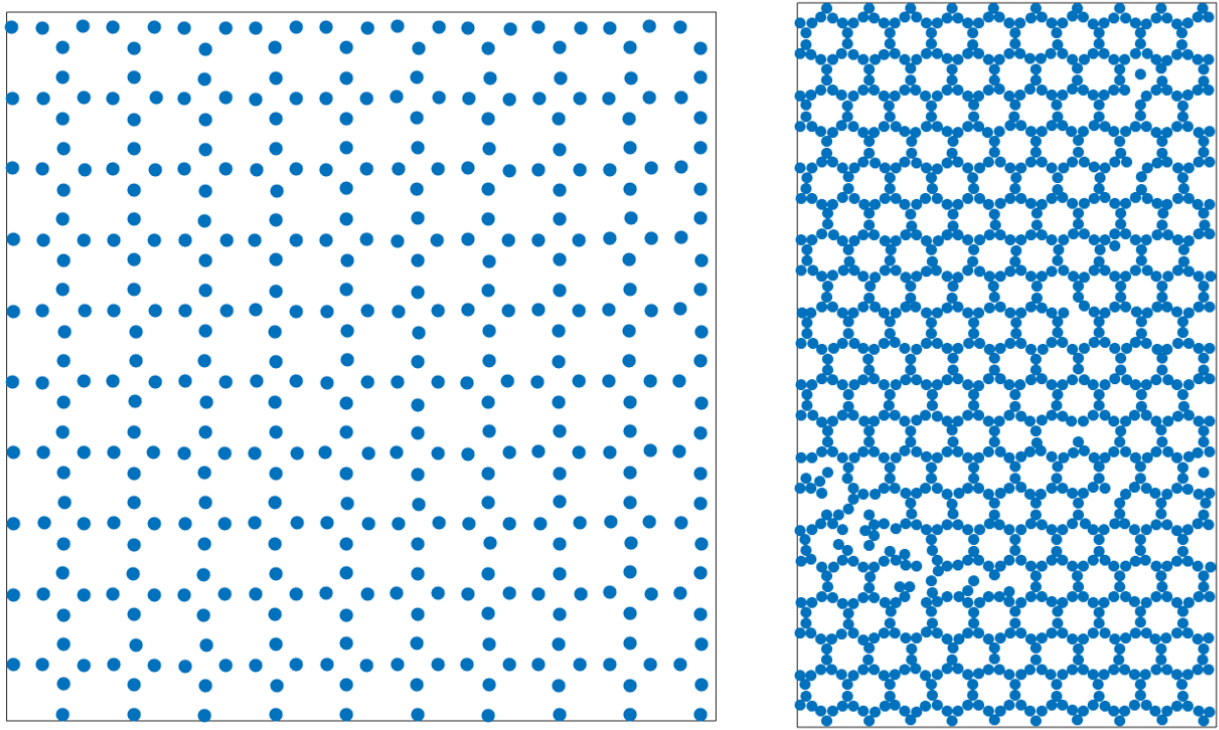}
	\caption{Configuration snapshots from molecular dynamics simulated annealing runs for systems of particles interacting through the RE optimized pair potentials of Fig.~\ref{fig:truncs_pots} for truncated square (left) and truncated hexagonal (right) target lattices.}
	\label{fig:truncs_sa}
	\end{figure} 
	
	Our designed single-well interactions for these target structures are displayed in figure \ref{fig:truncs_pots}. Indeed, as can be seen, both potentials display the expected features as per the applied `design rules' and, save for the single well, closely resemble those published for the same targets in previous RE work (figure $[3]$ \cite{RelEntropy_2D_structures}). The resulting self-assembled structures using the optimized interactions are shown in figure \ref{fig:truncs_sa}. Note that when we constrained the attractive well to be centered in a shell closer to the origin but still spanning the minimum interaction range (i.e. a broad well encompassing more than the outermost coordination shell(s)), the design strategy failed to stabilize any of these target structures, which further confirms the validity of design rule 2. 
	 . 

\section{Conclusions}
\label{sec:Conclusion}
Isotropic pair interactions comprising a repulsive core and a single attractive well are models for effective interactions that are ubiquitous in colloidal fluids and interesting for material design applications due to their possible tunability. Using GS and RE inverse design methods, we have inferred a set of `design rules' that help to understand the properties of single-well pair potentials that can stabilize three distinct two-dimensional target lattices (square, honeycomb and kagome). These rules can be summarized as (1) the interaction range must span a minimum number of coordination shells to differentiate the target structure from its competitors, (2) the well must be relatively narrow and located toward the end of the minimally required interaction range, and (3) nontrivial repulsive features (e.g., shoulders) in the potential spanning specific coordination shells in the target and competing lattices are required to impose energetic advantages to the target structure. 

We further examined two challenging, low-coordinated target structures using the RE inverse design approach:  truncated square and truncated hexagonal lattices. The optimized pair potentials, which displayed features consistent with the aforementioned design rules, readily assembled into the target structures upon cooling in computer simulations. Given their applicability across a wide variety of target structures, potential functional forms, and design methods, we believe these rules will be equally applicable to other crystal design targets in 3D, where transferability of 2D results to 3D targets are known to apply in repulsive systems\cite{AvniDimTransfer} and greatly simplify the design process.

\section{Supplementary Material}
See supplementary material for an illustration of specialized honeycomb and kagome competitors, parameters of reported GS potentials as well as derivations of the binary phase competitor calculations and the well minimum position update scheme used in the RE method. 

\begin{acknowledgments}
T.M.T. acknowledges support of the Welch Foundation (F-1696) and the National Science Foundation (CBET-1720595). We also acknowledge the Texas Advanced Computing Center (TACC) at the University of Texas at Austin for providing computing resources used to obtain results presented in this paper.
\end{acknowledgments}

%
\end{document}